\documentclass[aps,pr,twocolumn,groupedaddress,notitlepage,
floatfix,superscriptaddress]{revtex4-2}

\pdfoutput=1 
\usepackage{graphicx,graphics,epsfig,subfigure,times,bm,bbm,amssymb,amsmath,amsthm,mathrsfs,MnSymbol} \usepackage{gensymb} \usepackage{amsfonts} \usepackage{float} \usepackage[matrix,frame,arrow]{xypic} \usepackage[pdfstartview=FitH]{hyperref} 
\usepackage{times}
\usepackage{float}
\usepackage{graphics}
\usepackage[T1]{fontenc}

\usepackage{braket}  
\usepackage{enumerate} \usepackage[normalem]{ulem}
\usepackage[usenames,dvipsnames]{xcolor} \usepackage{multirow} \usepackage{mathtools}
\usepackage{bbm} \usepackage{titletoc} 

\definecolor{orange}{rgb}{1,0.5,0} 
\hypersetup{ colorlinks=true,       
	linkcolor=red,          
	citecolor=blue,        
	filecolor=magenta,      
	urlcolor=blue,           
	runcolor=cyan }

\begin{document}
	\title{Confinement-Induced  Enhancement of Superconductivity in a Spin-$\frac{1}{2}$ Fermion Chain Coupled to a $\mathbb{Z}_2$ Lattice Gauge Field }
	
	\author{Zi-Yong~Ge}
	\affiliation{Theoretical Quantum Physics Laboratory, Cluster for Pioneering Research, RIKEN, Wako-shi, Saitama 351-0198, Japan}
	
	\author{Franco Nori}
	\affiliation{Theoretical Quantum Physics Laboratory, Cluster for Pioneering Research, RIKEN, Wako-shi, Saitama 351-0198, Japan}
	\affiliation{Center for Quantum Computing, RIKEN, Wako-shi, Saitama 351-0198, Japan}
	\affiliation{Department of Physics, University of Michigan, Ann Arbor, Michigan 48109-1040, USA}

    \date{\today}
	\begin{abstract} 
	We investigate a spin-$\frac{1}{2}$ fermion chain minimally coupled to a $\mathbb{Z}_2$ gauge field.
	In the sector of the gauge generator $\hat G_j =-1$, the model reduces to the Hubbard model with repulsive onsite interaction coupled 
	to a $\mathbb{Z}_2$ gauge field.
    We uncover how electric fields affect low-energy excitations by both analytical and numerical methods.
	In the half-filling case, despite electric fields, the system is still a Mott insulator, just like the Hubbard model.
	For hole-doped systems, holes are confined under nonzero electric fields, resulting in a hole-pair bound state.
	Furthermore, this bound state also significantly affects the superconductivity, which manifests itself
	in the emergence of attractive interactions between bond singlet Cooper pairs.
	Specifically, numerical results reveal that the dimension of the dominant superconducting order parameter becomes smaller when increasing the electric field,
	signaling an enhancement of the superconducting instability induced by lattice fermion confinement.
	The superconducting order can even be the dominant order of the system for suitable doping and large applied electric field.
	The confinement also induces a $\pi$ momentum for the dominant superconducting order parameter
	leading to a quasi-long-range pair density wave order. Our results provide insights for understanding unconventional superconductivity in $\mathbb{Z}_2$ LGTs
	and might be experimentally addressed in quantum simulators.
	\end{abstract}

	\maketitle
	
	\section{Introduction}
	
	Lattice gauge theories (LGTs)  were originally proposed to understand the confinement of quarks 
	and became a fundamental concept in high-energy physics~\cite{PhysRevD.10.2445}.
	In correlated electronic systems, due to strong quantum fluctuations, LGTs can also emerge leading to many exotic quantum phases~\cite{RevModPhys.51.659,PhysRevB.65.024504,Fradkin2013,KITAEV20032,RevModPhys.89.025003,Senthil1490}.
	Moreover, LGTs have also applied to understand high-$T_c$ superconductors, 
	where $U(1)$~\cite{PhysRevB.37.580}, $SU(2)$~\cite{PhysRevB.38.745,PhysRevB.38.2926} 
	and $\mathbb{Z}_2$~\cite{PhysRevB.62.7850} gauge theories have been introduced
	in doped Mott insulators~\cite{RevModPhys.63.1,RevModPhys.66.763,RevModPhys.78.17}.
	However, studying  LGTs coupled to dynamical matter fields, especially in the 2D case, is a quite challenging task for conventional methods.
	Explicitly, there is a sign problem  for quantum Monte Carlo methods~\cite{RevModPhys.67.279,RevModPhys.73.33,doi:10.1126/science.abg9299}, 
	and the density matrix renormalization group (DMRG)~\cite{PhysRevLett.69.2863,RevModPhys.77.259,SCHOLLWOCK201196}  method is difficult to extend to high-dimensional systems due to  large entanglement entropies.
	Recently, with the rapid development of quantum simulations~\cite{doi:10.1080/00018730701223200,RevModPhys.80.885,Buluta108,Buluta_2011, RevModPhys.86.153,Gross995,s41567-019-0733-z}, 
	studying LGTs in synthetic quantum many-body systems has become possible~\cite{PhysRevA.73.022328,PhysRevLett.109.125302,
		PhysRevLett.109.175302,PhysRevX.3.041018,PhysRevLett.111.110504,PhysRevLett.117.240504,
		PhysRevLett.118.070501,Barbieroeaav7444,
		ISI:000494944200024,ISI:000494944200023,ISI:000591335100019,PhysRevX.10.021057,PhysRevX.10.021041,PhysRevResearch.2.023015,Ge_2021, PRXQuantum.2.030334,PhysRevResearch.4.L022060,doi:10.1126/science.abl6277,arXiv:2203.08905,arXiv:2205.08541}.
	Quantum simulations  provide an alternative for understanding or solving  hard problems in LGTs.
	Meanwhile, quantum simulators are also an outstanding platform to study non-equilibrium dynamics of LGTs~\cite{doi:10.1080/00107514.2016.1151199,banuls2020simulating,PRXQuantum.3.010324}.

	The simplest example of LGTs is $\mathbb{Z}_2$ gauge theories~\cite{RevModPhys.51.659,KITAEV20032,PhysRevD.17.2637,PhysRevD.19.3682}.
	Motivated by recent quantum simulation experiments, 
	studying $\mathbb{Z}_2$ LGTs coupled to dynamical matter fields has attracted considerable interests~\cite{PhysRevB.89.165416,PhysRevLett.118.266601,PhysRevLett.119.176601,PhysRevB.96.205104,PhysRevB.97.245137,di2019resolution,PhysRevB.99.075103,
		PhysRevA.102.023718,PhysRevX.10.041007,PhysRevLett.124.120503,10.21468/SciPostPhys.10.6.148,PhysRevLett.127.167203,
		PhysRevResearch.3.023079,PhysRevA.103.053703,PhysRevB.105.075132,PRXQuantum.3.020345}.
	These works mainly focus on single-component matter fields.
	However, to relate to unconventional superconductors,  spin-$\frac{1}{2}$ (two-component) fermions should be considered as the matter field.
	Generally, there exists different low-energy physics
	between single- and multi-component fermions coupled to gauge fields.
	For instance, deconfined phases are absent in 2D single-component fermions coupled to $\mathbb{Z}_2$ LGTs,
	while they can emerge in  two- or multi-component cases~\cite{PhysRevX.6.041049,Gazit2017}.
	However, $\mathbb{Z}_2$ LGTs coupled to spin-$\frac{1}{2}$ fermions	are not yet fully understood even in 1D, 
	especially the superconducting order in the sector of the gauge generator $\hat G_j =-1$ 
    with doped holes~\cite{PhysRevB.62.7850,Gazit2017,Silvi2017finitedensityphase}.
	Moreover, it is still an open question how to realize this system in quantum simulators.

	In this work, we present systematic investigations of a spin-$\frac{1}{2}$ fermion chain coupled minimally to a $\mathbb{Z}_2$ LGT.
	To relate to the physics of Mott insulators and unconventional superconductors, we mainly consider low-energy excitations in the $\hat G_j =-1$ sector.
	Thus, the system is equivalent to the Hubbard model  coupled to a $\mathbb{Z}_2$ gauge field.
	The presence of an electric field term induces a nonvanishing expectation value of  electric field  and  a large fluctuation of gauge field.
	Therefore, strings formed between charges are made of stable electric fields leading to the confinement of lattice fermions.
	We first present a phenomenological analysis of the ground state and derive the effective Hamiltonian in the large electric field limit.
    The half-filling system is still a Mott insulator with the spin sector being an antiferromagnetic Heisenberg model.
	Hence, similar to the Hubbard model, there is a deconfined spinon excitation, although lattice fermions are confined.
	However, in hole-doped systems, the electric field can induce confinement of holes resulting in the emergence of hole-pair bound states.
	Moreover, the kinetic term of the hole pair can contribute to an attractive interaction between bond singlet Cooper pairs, 
	which is expected to enhance the superconductivity.

	We also implement DMRG~\cite{PhysRevLett.69.2863,RevModPhys.77.259,SCHOLLWOCK201196}  
	methods to  support the above analytical discussion.
    Numerical results demonstrate that lattice fermions are indeed always confined in half-filling systems,
    whereas they becomes deconfined in hole-doped systems when electric fields are absent.
    In addition, there  exists a deconfined spinon excitation in both half-filling and hole-doped cases, just like the Hubbard model.
    Regarding superconductivity, we find that bond singlet pairs are the dominant superconducting order parameter in hole-doped systems.
    Remarkably, the dimension of this order parameter becomes smaller when increasing an applied electric field, revealing that the confinement of holes can enhance superconductivity.
    Meanwhile, the superconducting order can be the dominant order of the system for large applied electric field and suitable doping.
    Numerical results also show that the confinement can result in a $\pi$ momentum for the dominant superconducting order.
    Thus, there  exists a quasi-long-range pair density wave (PDW) correlation~\cite{Berg_2009,PhysRevB.79.064515,PhysRevLett.105.146403,PhysRevB.85.035104,PhysRevX.4.031017,RevModPhys.87.457}.
    Finally, we also propose an approach to implement our model in experimental quantum simulators.

  The rest of this paper is organized as follows. In Sec. \ref{Sec2}, we introduce the model of a 1D spin-$\frac{1}{2}$ fermion chain coupled to a $\mathbb{Z}_2$ gauge field. In Sec. \ref{Sec3}, we present a phenomenological discussion about the confinement of lattice fermions and the existence of hole-pair bound states. 
  In Sec. \ref{Sec4}, we derive the effective Hamiltonian of the system with large electric fields, and analyze how hole-pair bound states enhance the superconducting order.
  In Sec. \ref{Sec5}, we present numerical results calculated by DMRG methods to support the above discussion.
  A possible experimental implementation of this system is proposed  in Sec. \ref{Sec6}.
  Finally, in Sec. \ref{Sec7}, we summarize the results and give an outlook of our work.

	\section{Model}\label{Sec2}
	
	Here we consider a  spin-$\frac{1}{2}$ fermion chain coupled to a dynamical $\mathbb{Z}_2$ gauge field.
	The  Hamiltonian reads
	\begin{align} \label{H} \nonumber
		\hat  H = &-t\sum_{j=1}^{L-1}\sum_{\sigma=\uparrow,\downarrow}\big(\hat f^\dagger_{j,\sigma}\hat \tau^z_{j+\frac{1}{2}}\hat f_{j+1,\sigma} + \text{H.c.}\big)\\
		&-h\sum_{j=1}^L \hat \tau_{j+\frac{1}{2}}^x-\frac{U}{4}\sum_{j=1}^{L-1}\hat \tau_{j-\frac{1}{2}}^x\hat \tau_{j+\frac{1}{2}}^x,
	\end{align}
	where $\hat f^\dagger_{i,\sigma}$ ($\hat f_{j,\sigma}$) is the creation (annihilation) operator of the fermion living on site $j$,
	$\hat \tau_{j+\frac{1}{2}}^\alpha$ is the Pauli matrix acting on the link between sites $j$ and $(j+1)$ [labeled by $(j+\frac{1}{2})$],
	and $L$ is the system size.
	The first term describes fermions coupled minimally to gauge fields via the Ising version of the Peierls substitution with amplitude $t>0$. 
	The second term is an electric field with strength $h>0$.
	The third term is a ferromagnetic Ising interaction of electric fields with strength $U>0$.

	The Hamiltonian $\hat H$ is $\mathbb{Z}_2$ gauge invariant with a generator defined as 
	\begin{align} 
		\hat G_j = \hat \tau_{j-\frac{1}{2}}^x (-1) ^{\hat N_j}\hat\tau_{j+\frac{1}{2}}^x,
	\end{align}
	where $\hat N_j = \hat n_{j,\uparrow}+\hat n_{j,\downarrow}$ ($\hat n_{j,\sigma} = \hat f^\dagger_{j,\sigma}\hat f_{j,\sigma}$) is the fermion number on site $j$.
	In addition to gauge invariance, $\hat H$ also possesses a global $SU_{\text{s}}(2)\times U(1)$ symmetry for the arbitrary filling factor.
	Here $U(1)$ represents the conservation of total fermion number, i.e., $[\sum_j\hat N_j,\hat H]=0$,
	and $SU_{\text{s}}(2)$  corresponds  the spin rotation.
	Specifically, for the half filling, due to the particle-hole symmetry, this continuous symmetry is enlarged to
	$SU_{\text{s}}(2)\times SU_{\eta}(2)\times U(1)$,
	where $SU_{\eta}(2)$ is the pseudospin rotation symmetry~\cite{doi:10.1142/S0217984990000933,PhysRevLett.65.120}.

	The 2D version of Eq.~(\ref{H}) in the $\hat G_j =-1$ sector is potentially relevant for understanding
	superconductivity in doped Mott insulators~\cite{PhysRevB.62.7850}. 
	However, this is challenging for conventional numerical and analytic methods~\cite{PhysRevX.6.041049,Gazit2017}.
	In the following, we will demonstrate that the 1D case can be addressed analytically by perturbation theory in the strong electric field limit.
	To relate to the physics of Mott insulators and superconductivity, we focus on the $\hat G_j =-1$ gauge sector.
	Thus, the Ising interaction of gauge fields under this gauge constraint is equivalent to an onsite repulsive interaction of fermions
	  \begin{align} 
		-\hat \tau_{j-\frac{1}{2}}^x\hat \tau_{j+\frac{1}{2}}^x=(1-2\hat n_{j,\uparrow})(1-2\hat n_{j,\downarrow}).
	\end{align}
	Therefore, the Hamiltonian (\ref{H})  reduces to a Hubbard model when $h=0$.
	Hereafter, we mainly study how electric fields affect the low-energy physics of the Hubbard model.

   \section{Charge configurations of ground states}\label{Sec3}
   
   To shed light on how the confinement of lattice fermions occurs,
   we first  analyze phenomenologically  the configuration of the charge sector in ground states.
   For simplicity, we can consider the limit $h,U\gg t$,
   where the terms 
    \begin{align} 
    \hat H_h:=-h\sum_{j=1}^L \hat \tau_{j+\frac{1}{2}}^x,
   \end{align}
   and
   \begin{align}
   \hat H_U:=-\frac{U}{4}\sum_{j=1}^{L-1}\hat \tau_{j-\frac{1}{2}}^x\hat \tau_{j+\frac{1}{2}}^x,
   \end{align}
  dominate the energy.

   In the half-filling case, if each site occupies one fermion, then under the gauge constraint  $\hat G_j=-1$, 
   all $\tau$ spins can be polarized at $\braket{\hat\tau^x_{j+\frac{1}{2}}}=1$, simultaneously.
   This configuration can indeed minimize the energy of both $\hat H_h$ and $\hat H_U$.
   In addition, exciting a double occupation and a hole, which can be considered as a meson, at least costs energy $\sim (h+U)$,
   i.e., there is a charge gap.
   Thus, the system is still a Mott insulator identical to the Hubbard model, and lattice fermions $\hat f_{j,\sigma}$ should be confined.
   Here, the existence of electric fields can only enlarge the charge gap, but cannot change the Mott insulator phase in this case.
   We note that the situation becomes different when $U<0$, describing antiferromagnetic interactions of electric fields
   	(or attractive on-site interactions of fermions) in the gauge sector $\hat G_j = -1$. 
   	
   	In Appendix~\ref{A3}, we present a detailed discussion for $U<0$. 
   We find that there exists a quantum phase transition from the Mott insulator (gapless) to a meson condensed phase (gaped) when increasing $-U$.
   Here, the meson condensed phase  spontaneously breaks the translational symmetry, and the corresponding matter field is a pseudo-spin valence-bond solid.

   In hole-doped systems, it is not possible for each site to be occupied by a fermion,
   so not all $\tau$ spins can be polarized at $\braket{\hat\tau^x_{j+\frac{1}{2}}}=1$ in the $\hat G_j =-1$ sector.
   In addition, double occupation should also be suppressed due to its energy cost $\sim (h+U)$. 
   Thus, only holes and single-occupation states are allowed in the ground state.
   To reveal the configuration of the charge sector,
   we take the case of two holes as an example (to fix the  even fermion parity subspace, we only consider an even number of holes.).
   The corresponding gauge-invariant configuration has the form 
   \begin{align} \label{gs} 
   	\ket{...+1+0-1-1-...-1-1-0+1+1+...},
   \end{align}
   where $\ket{\pm}$ labels the eigenstate of $\hat \tau^x$ with  eigenvalue $\pm1$, 
   and $\ket{0}$ ($\ket{1}$) labels a hole (single occupation).
   In Eq.~(\ref{gs}), there are $r$ links polarized at the state $\ket{-}$, where $r$ is the distance between two holes.
   Thus, there is a string tension between two holes with energy $hr$,
   so the hole becomes confined when $h\neq0$.
   For large $h$, to minimize the energy, these two holes should be bonded on two nest-neighbor (NN) sites.
   Generalizing to the multi-hole systems, we can find that holes must exist in pairs, i.e., form hole-pair bound states $\ket{0-0}$, 
   which are absent in the Hubbard model.
   Therefore, in the presence of doped holes, the existence of $\hat H_h$ can significantly affect the charge degrees of freedom,
   resulting in a distinct low-energy physics from the Hubbard model.

	\begin{figure}[t] \includegraphics[width=0.45\textwidth]{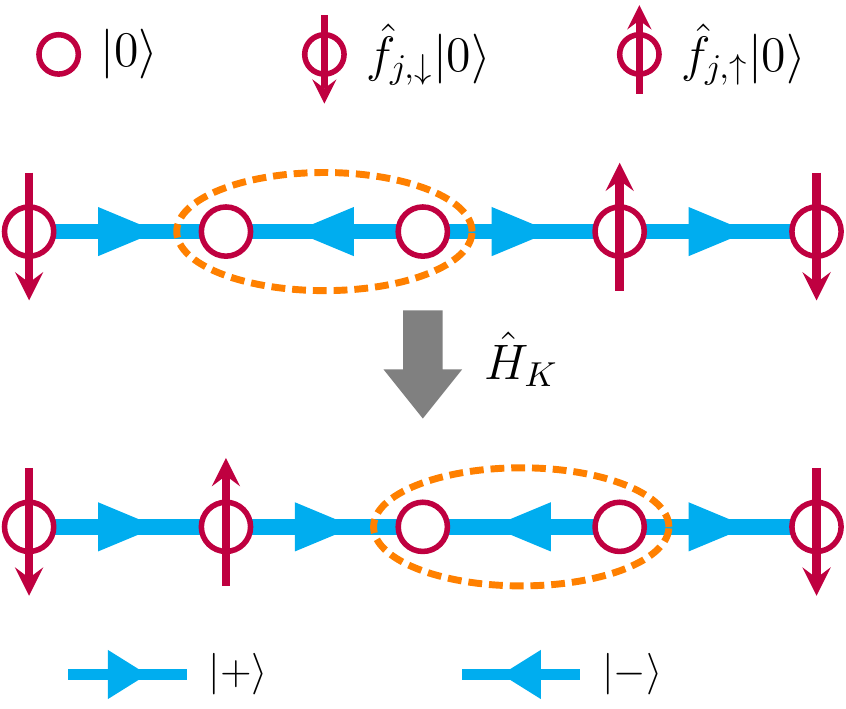} 
		\caption{Diagram of the hole pair dynamics. 
			The orange dashed ellipses represents a hole-pair bound state $\ket{0-0}$, which can hop to  NN bonds under the action of $\hat H_K$.  }
		\label{fig_1}
	\end{figure}
	
	\section{Effective Hamiltonian in the strong electric field limit}\label{Sec4}
	
	To further uncover the effect of electric fields, 
	we can consider the effective Hamiltonian in the limit of strong electric field $h\gg t$ and $U\rightarrow0$.
	Using the Schrieffer-Wolf transformation~\cite{PhysRev.149.491,BRAVYI20112793}, we can obtain 
	\begin{align} \label{Heff}\nonumber
		       &\hat  H_{\text{eff}} = \hat H_h + \hat H_S + \hat  H_{K}\\ \nonumber
				&\hat  H_{S} =J\sum_{j}\hat \tau^x_{j+\frac{1}{2}}\cdot\big( \hat{\boldsymbol{s}}_j\cdot\hat{\boldsymbol{s}}_{j+1} 
				+ \hat{\boldsymbol{\eta}}_j\cdot\hat{\boldsymbol{\eta}}_{j+1} \big), \\ 
				&\hat  H_{K} \!=\!\frac{J}{2}\sum_{j,\sigma}\hat \tau^x_{j+\frac{1}{2}}\!\cdot \!\big(\!\hat \tau^z_{j-\frac{1}{2}}\hat \tau^z_{j+\frac{1}{2}}\!+\!\hat \tau^y_{j-\frac{1}{2}}\hat \tau^y_{j+\frac{1}{2}}\!\big)\big(\!\hat f^\dagger_{j-1,\sigma}\hat f_{j+1,\sigma} \!-\! \text{H.c.}\big),
	\end{align}
	where the effective coupling $J=t^2/h$. In addition, $\hat{\boldsymbol{s}}_{j}=(\hat s^x_j,\hat s^y_j,\hat s^z_j)$  is the spin operator defined as 
	\begin{align}
		\hat{\boldsymbol{s}}_{j} := \sum_{\alpha,\beta}\hat f^\dagger_{j,\alpha} {\boldsymbol{\sigma}}_{\alpha\beta} \hat f_{j,\beta},
	\end{align}
   where ${\boldsymbol{\sigma}}$ are Pauli matrices,
    and $\hat{\boldsymbol{\eta}}_j = (\hat \eta^x_j,\hat \eta^y_j,\hat \eta^z_j)$ is the pseudospin operator generated by~\cite{doi:10.1142/S0217984990000933,PhysRevLett.65.120}
		\begin{align} \label{eta}
			\hat \eta^+_j =\hat \eta^x_j+i \hat \eta^y_j = (-1)^j\hat f^\dagger_{j,\uparrow} \hat f^\dagger_{j,\downarrow}.
	   \end{align}
Detailed derivations of Hamiltonian (\ref{Heff}) are presented in Appendix~\ref{A1}.

In the case of half filling, as discussed in the Sec.~\ref{Sec3},  all links are nearly polarized at $\braket{\hat\tau^x_{j+\frac{1}{2}}}=1$, and each site occupies one fermion.
Thus we have
\begin{align} 
\hat{\boldsymbol{\eta}}_j\cdot\hat{\boldsymbol{\eta}}_{j+1}=\hat \tau^z_{j-\frac{1}{2}}\hat \tau^z_{j+\frac{1}{2}}\!+\!\hat \tau^y_{j-\frac{1}{2}}\hat \tau^y_{j+\frac{1}{2}}=0.
\end{align}	
The effective Hamiltonian in this case is reduced to a 1D antiferromagnetic Heisenberg model (the constant is neglected)
\begin{align} 
	\hat  H_{\text{eff}}^{\text{hf}} = J\sum_{j}\hat{\boldsymbol{s}}_j\cdot\hat{\boldsymbol{s}}_{j+1}.
\end{align}	
This  is identical to the Hubbard model with large repulsive on-site interactions.
Therefore, in addition to the charge sector,
the electric field cannot affect the physics of the spin sector for the half filling.
According to properties of the 1D Heisenberg model~\cite{RevModPhys.63.1,Giamarchi2004,Fradkin2013},
there will be a deconfined spinon excitation.
The spin operator $\hat{\boldsymbol{s}}_{j}$ is indeed gauge invariant, so this spinon excitation is physically allowed.

For hole-doped systems, since double occupations are still forbidden, 
according to Eq.~(\ref{eta}), the term $\hat{\boldsymbol{\eta}}_j\cdot\hat{\boldsymbol{\eta}}_{j+1}$ only contributes a density-density interaction.
The Heisenberg term $\hat{\boldsymbol{s}}_j\cdot\hat{\boldsymbol{s}}_{j+1}$ cannot vanish, 
if and only if the NN sites $j$ and $(j+1)$ are both occupied by a single fermion, which corresponds to the link $(j+\frac{1}{2})$ for the state $\ket{+}$.
Thus, $\hat \tau^x_{j+\frac{1}{2}}\cdot\big( \hat{\boldsymbol{s}}_j\cdot\hat{\boldsymbol{s}}_{j+1} \big) $ is equivalent to $ \hat{\boldsymbol{s}}_j\cdot\hat{\boldsymbol{s}}_{j+1}$,
which is also gauge invariant.
Thus, the effective Hamiltonian can be written as
\begin{align}\nonumber
	&\hat  H_{\text{eff}}^{\text{hd}} = J\sum_{j}\big( \hat{\boldsymbol{s}}_j\cdot\hat{\boldsymbol{s}}_{j+1} + \hat N_j \hat N_{j+1}\big) \\
	&+\frac{J}{2}\sum_{j,\sigma}\hat \tau^x_{j+\frac{1}{2}}\cdot \big(\hat \tau^z_{j-\frac{1}{2}}\hat \tau^z_{j+\frac{1}{2}}+\hat \tau^y_{j-\frac{1}{2}}\hat \tau^y_{j+\frac{1}{2}}\big)\big(\hat f^\dagger_{j-1,\sigma}\hat f_{j+1,\sigma} - \text{H.c.}\big).
\end{align}	
The second term in $\hat  H_{\text{eff}}^{\text{hd}}$ is the next-nearest-neighbor (NNN) hoping term of holes.
If the site $j$ is a single-occupation state, then $\braket{\hat \tau^x_{j-\frac{1}{2}}}=\braket{\hat \tau^x_{j+\frac{1}{2}}}=1$,
leading to  $\hat \tau^z_{j-\frac{1}{2}}\hat \tau^z_{j+\frac{1}{2}}+\hat \tau^y_{j-\frac{1}{2}}\hat \tau^y_{j+\frac{1}{2}}=0$.
If the site $j$ is a hole, then $\braket{\hat \tau^x_{j-\frac{1}{2}}}=-\braket{\hat \tau^x_{j+\frac{1}{2}}}$
and  $\hat \tau^z_{j-\frac{1}{2}}\hat \tau^z_{j+\frac{1}{2}}=\hat \tau^y_{j-\frac{1}{2}}\hat \tau^y_{j+\frac{1}{2}}\neq0$.
Thus, the NNN hoping of holes between $(j-1)$ and $(j+1)$ can contribute, if and only if the site in between is also a hole.
That is, this term allows a hole-pair bound state $\ket{0-0}$ to hop to NN sites [see Fig.~\ref{fig_1}].
Therefore, in the hole-doped system, the dynamics of the charge sector is contributed by hole pairs, while the spin sector is still an antiferromagnetic Heisenberg model.

	   		\begin{figure}[t] 
		\includegraphics[width=0.5\textwidth]{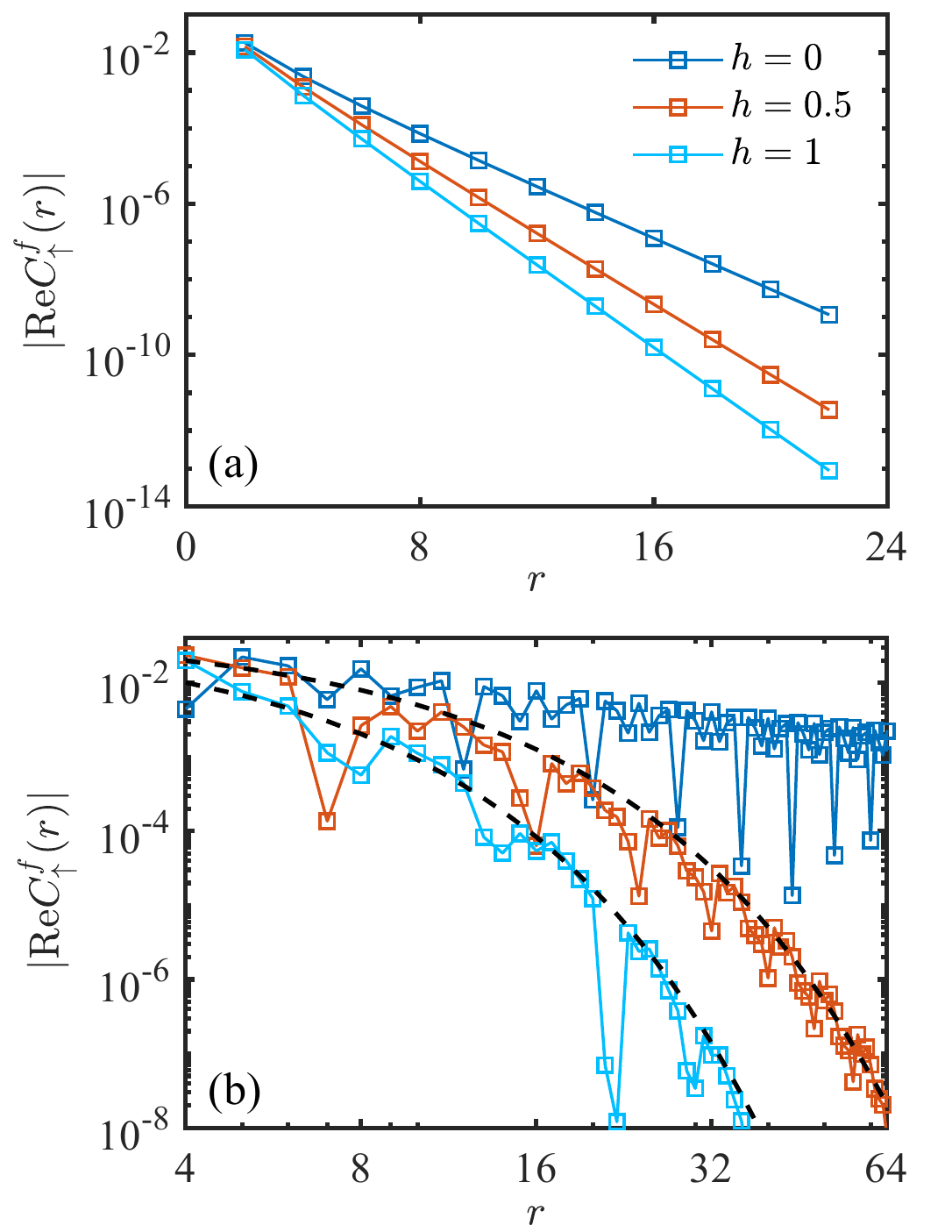} 
		\caption{Correlation functions of lattice fermions  $C^{\text{f}}_\uparrow (r)$ for (a) half and (b) $3/8$ filling, respectively.
			Here we choose $L=256$ and $U=8$.
			The black dashed curves are fitting exponential functions.
			To reduce  finite-size effects, we calculate the correlation function between sites $L/2$ and $L/2+r$.}
		\label{fig_2}
	\end{figure}
	
	\begin{figure}[t] 
		\includegraphics[width=0.5\textwidth]{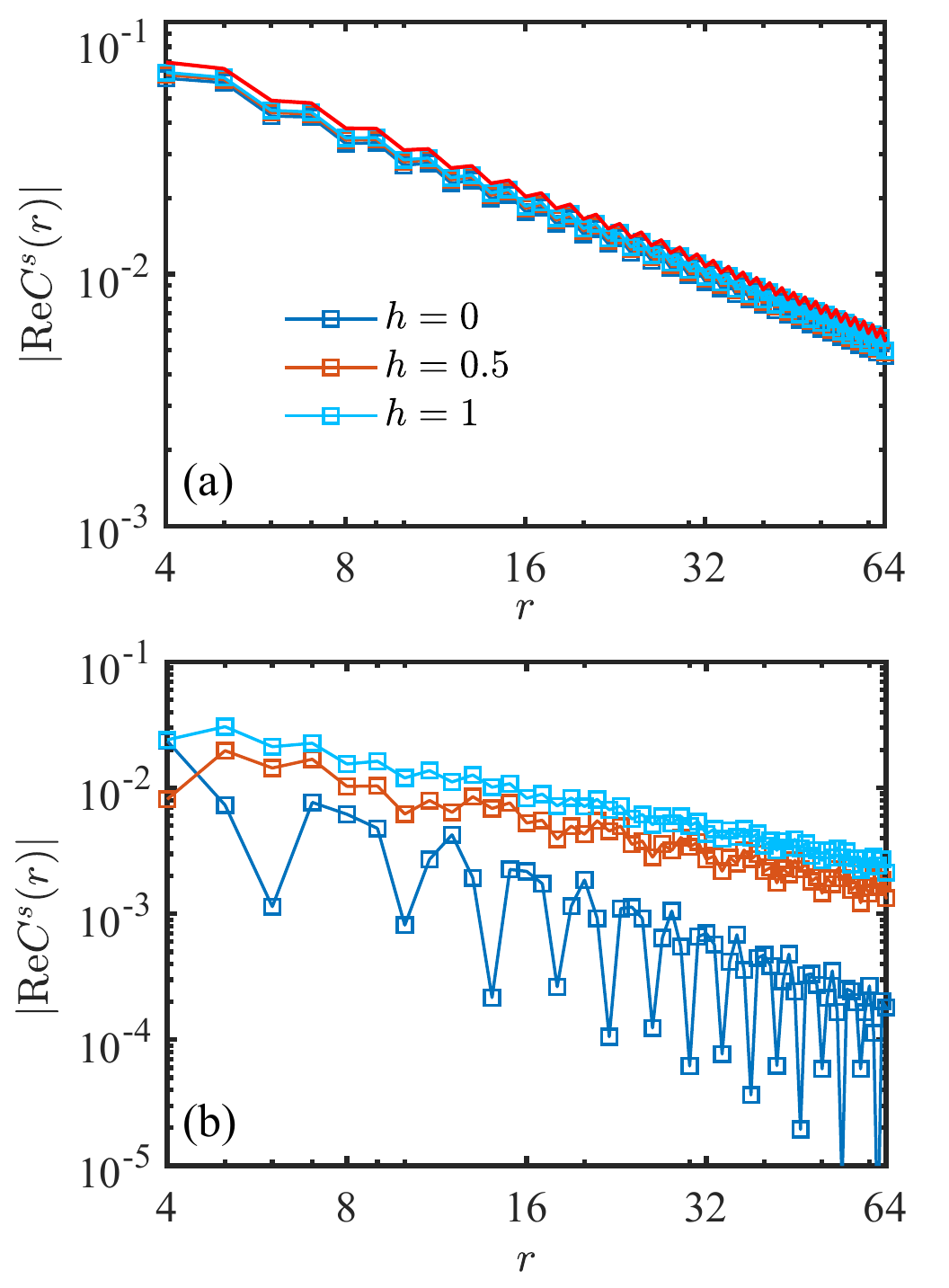} 
		\caption{Correlation functions of spins $C^{\text{s}} (r)$ for (a) half and (b) $3/8$ filling, respectively. 
			Here we choose $L=256$ and $U=8$.
			The red line is the spin-exchange correlation function of 1D Heisenberg model.
			To reduce  finite-size effects, we calculate the correlation function between sites $L/2$ and $L/2+r$.}
		\label{fig_3}
	\end{figure}

	\begin{figure*}[t] 
		\includegraphics[width=1\textwidth]{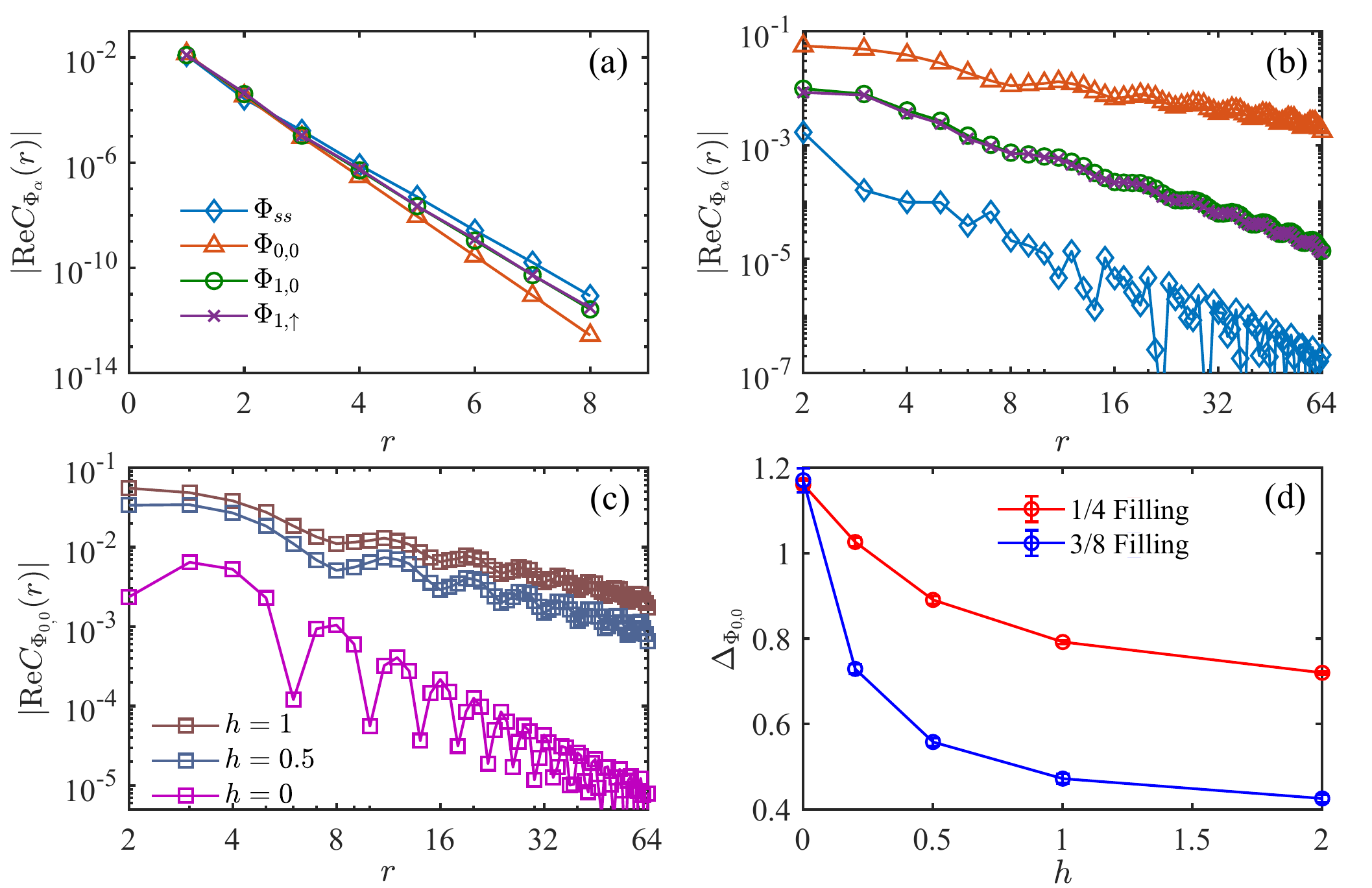} 
		\caption{Correlation functions of superconducting order parameters. 
			The scaling of different superconducting order parameters with $h=1$, $U=8$ and $L=256$ in the cases of 
			(a) half filling and (b) $3/8$ filling.
			(c) Correlation functions of bond singlet pairs for different electric fields.
			(d) The dimension of  bond singlet pairs versus electric fields.
			To reduce finite-size effects, we calculate the correlation function $C_{\Phi_\alpha}(r)=\braket{\hat \Phi^\dagger_\alpha (L/2+r) \hat \Phi_\alpha (L/2)}$.}
		\label{fig_4}
	\end{figure*}

	We further analyze the kinetic term $\hat  H_{K}$
	to uncover how the hole-pair bound state affects the superconductivity.
	Here, $\braket{\hat \tau^x_{j-\frac{1}{2}}}=-\braket{\hat \tau^x_{j+\frac{1}{2}}}$, if site $j$ is a hole, 
	and $\braket{\hat \tau^x_{j+\frac{1}{2}}}=\pm1$ with a hole pair at the link $(j\pm\frac{1}{2})$.
	Thus, $H_K$ can be simplified as 
	\begin{align}\label{hk2}
		\hat  H_{K} = J\sum_{j,\sigma} \hat f^\dagger_{j-1,\sigma}\hat \tau^z_{j-\frac{1}{2}}\hat \tau^z_{j+\frac{1}{2}}\hat f_{j+1,\sigma} + \text{H.c.}.
	\end{align}	
	In addition, since the site $j$ should be a hole, we can insert a term $\hat f_{j,\sigma}\hat f^\dagger_{j,\sigma}$ into Eq.~(\ref{hk2}).
	Therefore, this term can be rewritten as interactions of bond Cooper pairs
    \begin{align} \label{cpK}\nonumber
	\hat  H_{K}&\! =\! J\sum_{j} \hat \Phi_{1,0} (\! j-\!1) \hat \Phi^\dagger_{1,0} (j )-\hat \Phi_{0,0} (j-\!1) \hat \Phi^\dagger_{0,0} (j) + \text{H.c.}\\
     &\!=J\sum_{j} \hat \Phi_{1,\uparrow} (\! j - \! 1) \hat \Phi^\dagger_{1,\uparrow} (j)+\hat \Phi_{1,\downarrow} (\! j-\! 1) \hat \Phi^\dagger_{1,\downarrow} (j) + \text{H.c.},
	\end{align}
   where $\hat \Phi^\dagger_{s,m}$ ($s=0,1$ and $m=0,\uparrow,\downarrow$) is the gauge-invariant order parameter of bond Cooper pairs with the form
   \begin{align} \label{Bcooppa}\nonumber
   	&\hat \Phi^\dagger_{0,0} (j) := \frac{1}{\sqrt{2}}\hat \tau^z_{j+\frac{1}{2}}\cdot(\hat f^\dagger_{j,\uparrow}\hat f^\dagger_{j+1,\downarrow}-\hat f^\dagger_{j,\downarrow}\hat f^\dagger_{j+1,\uparrow}),\\ \nonumber
   	&\hat \Phi^\dagger_{1,0} (j) := \frac{1}{\sqrt{2}}\hat \tau^z_{j+\frac{1}{2}}\cdot(\hat f^\dagger_{j,\uparrow}\hat f^\dagger_{j+1,\downarrow}+\hat f^\dagger_{j,\downarrow}\hat f^\dagger_{j+1,\uparrow}),\\
   	&\hat \Phi^\dagger_{1,\sigma} (j) :=\hat f^\dagger_{j,\sigma}  \hat \tau^z_{j+\frac{1}{2}} \hat f^\dagger_{j+1,\sigma}.
   \end{align}
    Here $\hat \Phi^\dagger_{0,0}$ represents a singlet pair, while the other three ones are triplet pairs.
   In an antiferromagnetic background, triplet pairing should be suppressed due to a large energy cost.
   Meanwhile, they also exhibit repulsive interactions in Eq.~(\ref{cpK}), which are generally irrelevant to the superconductivity.
   However, the situation is different for singlet pairs, which are low-energy pairings in antiferromagnetic backgrounds 
   and have attractive interactions in $\hat H_K$.
   Therefore, $\hat \Phi_{0,0} $ should be the dominant pairing that contributes to the charge dynamics.
   Moreover, this term should be also expected to enhance the superconducting instability due to the attractive interaction.
  Thus we expect a superconducting single phase under large electric fields and proper doping.

	\section{Numerical simulations.}\label{Sec5}
	To verify the above results, we need to perform  numerical simulations.
	We implement  the DMRG algorithm, which is one of the most efficient methods to numerically study 1D quantum many-body systems.
	We project ground states to the specific gauge sector by adding a Lagrange multiplier $\lambda_j$ to the original Hamiltonian, i.e.,
	calculating the Hamiltonian 
	\begin{align}
	\hat H_{\text{num}}:=\hat H + \sum_j \lambda_j\hat G_j.
	\end{align}
	Since $[\hat G_j, \hat H]=0$, $\hat H$ should have the same eigenstates as $\hat H_{\text{num}}$.
	When $\lambda_j\gg 1$, the ground state of $\hat H_{\text{num}}$ satisfies $\braket{\hat G_j}=-1$,
	which is  the ground state of $\hat H$ in the {\color{red}$\hat G_j =-1$} sector.
	Here we mainly study how electric fields affect excitations of this system in the presence of large $U$,
	i.e., the effect of confinement concerning the 1D Hubbard model.
	To probe excitations, we need to calculate the corresponding gauge-invariant correlation functions.
	During the calculation, we choose $\lambda_j = 100$, $t=1$, and open boundary conditions with system size up to $L=256$.
	The maximum bond dimension is $\chi=600$ and truncation errors $\sim10^{-7}$.
	The expectation value of the $\mathbb{Z}_2$ gauge generator satisfies $|\braket{\hat G_j}+1|<10^{-12}$.
	We note that, despite the absence of rigorous gauge invariance, this numerical method can naturally extend to study experimental imperfections.

\begin{figure*}[t] \includegraphics[width=1\textwidth]{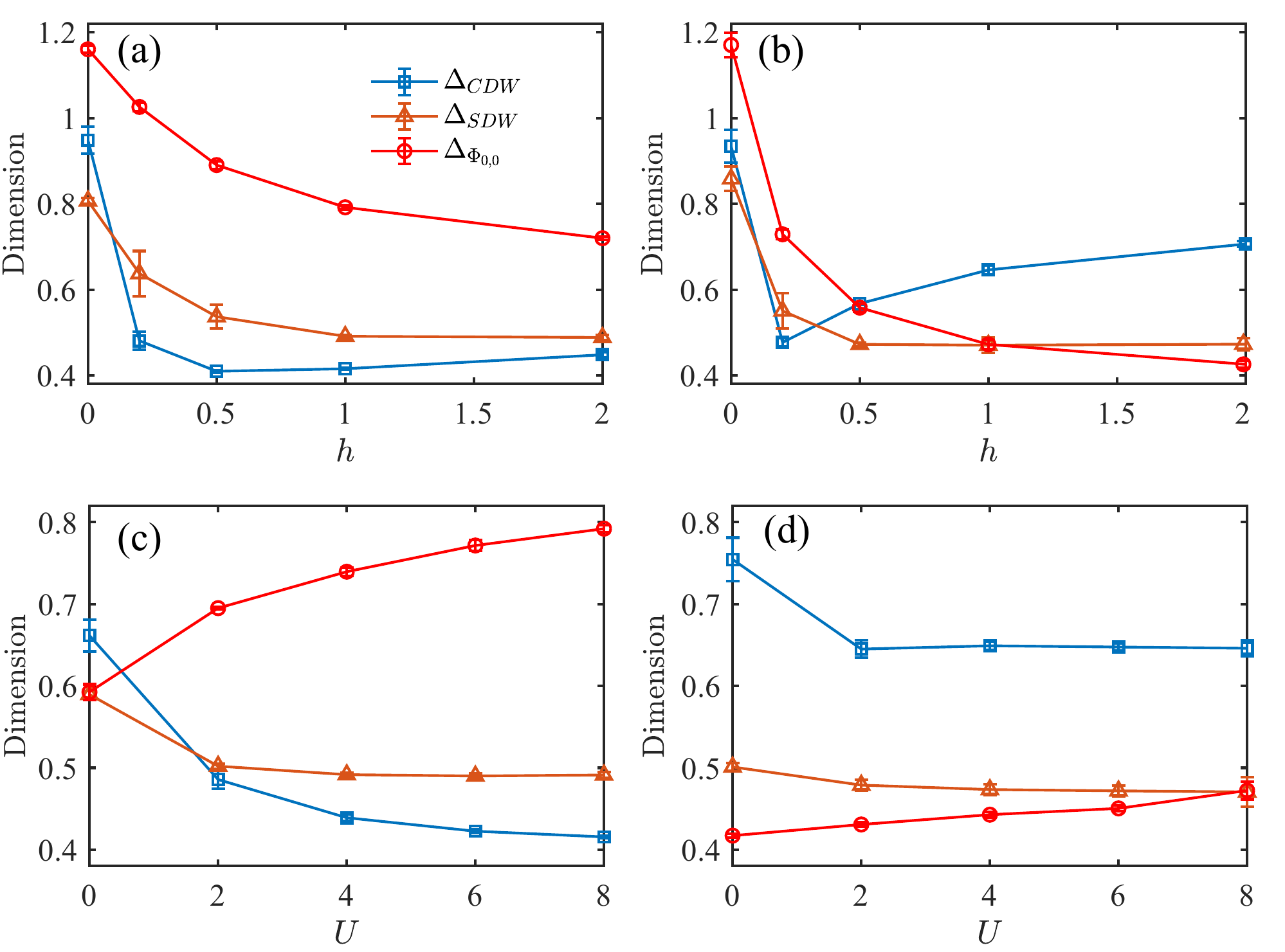} 
	\caption{Dimensions of CDW, SDW and  bond-singlet pairs. Dimensions of these order parameters versus  applied electric field $h$ for (a) 1/4 filling and (b) $3/8$ filling, respectively.
			We fix $U=8$. Dimensions versus $U$ for (c) 1/4 filling and (d) $3/8$ filling. We fix $h=1$.}
	\label{fig_5}
\end{figure*}
	
	\subsection{Lattice fermions}
	We first consider the confinement and deconfinement of lattice fermions by introducing a string correlation function~\cite{PhysRevLett.124.120503}
		\begin{align}\label{cf}
		C^{\text{f}}_\sigma (i-j) := \braket{\hat f^\dagger_{i,\sigma}\big(\prod_{i\leq\ell <j}\hat \tau^z_{\ell+\frac{1}{2}}\big)\hat f_{j,\sigma}}.
	\end{align}
	As shown in Fig.~\ref{fig_2}(a), for the large $U$, $C^{\text{f}}_\sigma (i-j)$ shows exponential decay in the half-filling case
	for arbitrary electric fields (including $h=0$).
	Thus, the lattice fermion is always confined for finite $U$, which is consistent with the Mott insulator in this case.
	For hole-doped systems, when $h=0$, $C^{\text{f}}_\sigma$ shows a power law decay, 
	indicating a deconfinement of $\hat f^\dagger_{i,\sigma}$ [see Fig.~\ref{fig_2}(b)].
	However, it becomes confined for finite $h$, which is consistent with the string tension between holes.

   \subsection{Spin sector}
	For the spinon excitation, we define a spin-exchange correlation function
	\begin{align}
		C^{\text{s}} (i-j):= \braket{\hat s^+_i \hat s^-_j }.
	\end{align}
	which is also gauge invariant.
    Figures~\ref{fig_3}(a--b) show that this correlation function has a power-law scaling in both half-filling and hole-doped systems,
    indicating the existence of deconfined spinon excitations.
    Moreover, in the half-filling case, Fig.~\ref{fig_3}(a) shows that the $C^{\text{s}}$ is nearly identical to the spin-exchange correlation function of the 1D Heisenberg model for arbitrary $h$, indicating that the increase in $h$ can hardly affect the effective Hamiltonian (Heisenberg model).

    \subsection{Superconductivity}
    
	Now we focus on the superconductivity in hole-doped systems.
	For 1D spin-$\frac{1}{2}$ fermions, in addition to bond pairs defined in Eq.~(\ref{Bcooppa}), 
	we can also introduce a site singlet pair $\hat \Phi^\dagger_{\text{ss}} (j) := \hat f^\dagger_{j,\uparrow}\hat f^\dagger_{j,\downarrow}$.
	To study the superconductivity, we need to calculate correlators of the above Cooper pair order parameters, i.e.,
		\begin{align}
	C_{\Phi_\alpha}(i-j) := \braket{\hat \Phi^\dagger_\alpha (i) \hat \Phi_\alpha (j)},
	\end{align}
	where $\alpha = \text{ss}/0,0/1,0/1,\uparrow/1,\downarrow$.
	As shown in Fig.~\ref{fig_4}(a), $C_{\Phi_\alpha}$ exhibits an exponential decay for all superconducting order parameters in the case of half-filling
	suggesting the Mott insulator.
	In Fig.~\ref{fig_4}(b), we present the result of $C_{\Phi_\alpha}$ with $3/8$ filling.
	It shows that all types of Cooper pairs have power law scalings 
	\begin{align}
		C_{\Phi_\alpha}(i-j)\sim |i-j|^{-2\Delta_{\Phi_\alpha}},
	\end{align}
	where $\Delta_{\Phi_\alpha}$ is the dimension of $\hat \Phi_\alpha$.
	In addition, the bond singlet pair is indeed the dominant pairing, which is consistent with Eq.~(\ref{cpK}).
	Moreover, the dimension of  this pair becomes smaller when increasing $h$ [see Figs.~\ref{fig_4}(c,d)].
	Thus, the electric field can indeed  enhance superconductivity, 
	which is consistent with the effective Hamiltonian in Eq.~(\ref{Heff}).
	      
	\begin{figure}[t] 
		\includegraphics[width=0.49\textwidth]{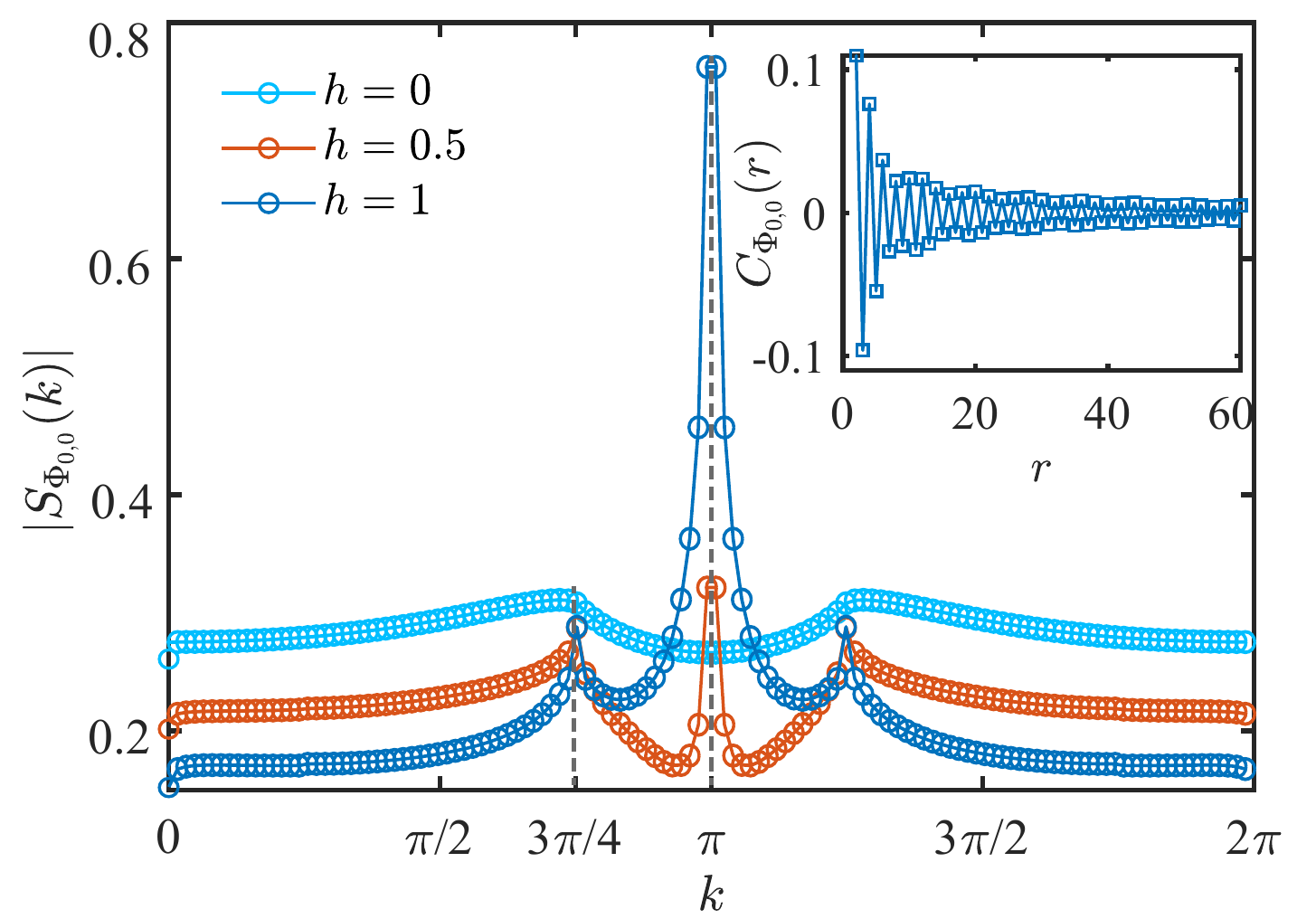} 
		\caption{Fourier analysis of the bond-singlet pair correlation for different $h$.
				We choose 3/8 filling and $L=256$. The insert is the corresponding correlation function in real space for $h=1$, 
				where there exists an oscillation with period of 2 sites.}
		\label{fig_6}
	\end{figure}

	   Now we discuss the dominant order in hole-doped systems.
	   In addition to the superconducting order, 
	   other two possible orders in interacting spin-$\frac{1}{2}$ fermion systems are charge density wave (CDW) and spin density wave (SDW).
	   Here, CDW order can be probed by the 
	   density-density correlation function
	   \begin{align} \label{Ccdw}
	   	C_{\text{N}} (i-j):= \braket{\hat N_i \hat N_j }- \braket{\hat N_i  } \braket{ \hat N_j },
	   \end{align}
	   and the SDW order can be probed by the spin correlation function
	   \begin{align} \label{Csdw}
	   	C_{\boldsymbol{s}} (i-j):= \braket{\hat{\boldsymbol{s}}_i \cdot \hat{\boldsymbol{s}}_j}.
	   \end{align}
      In hole-doped systems, similar to the $C_{\Phi_\alpha}$, they also exhibit power-law decay
       \begin{align} \nonumber
      &	C_{\text{N}} (i-j)\sim |i-j|^{-2\Delta_{\text{CDW}}}, \\   
      	&C_{\boldsymbol{s}} (i-j)\sim |i-j|^{-2\Delta_{\text{SDW}}}, 
      \end{align}
      where $\Delta_{\text{CDW}}$ and $\Delta_{\text{SDW}} $ are dimensions of the CDW and SDW, respectively.
      The most slowly decaying correlation function, i.e., the smallest dimension, means that the corresponding order parameter is dominant in the system.

      In Figs.~\ref{fig_5}(a--b), we present the dimensions of CDW, SDW and bond-singlet pairs versus electric fields for different filling factors.
      When $h=0$, i.e., the Hubbard model, we can find that $\Delta_{\text{CDW}}$ and $\Delta_{\text{SDW}} $ are both smaller than $\Delta_{\Phi_{0,0}}$,
      so the dominant order in a hole-doped system is not superconductivity.
      This is consistent with the result obtained by the bosonization method~\cite{Giamarchi2004,Fradkin2013}.
	  However, when increasing $h$, the situation becomes different.
	  For instance, in the case of $3/8$ filling and $h=2$,
      the dimension of bond-singlet pairs becomes the smallest one, see Fig.~\ref{fig_5}(b).	  
      Therefore, superconducting order can be dominant under the proper doping and large $h$.
       In Figs.~\ref{fig_5}(c--d), we present the dimensions of the above order parameters versus $U$ for different filling factors.
       The numerical result shows that the on-site repulsive interaction can enhance  the CDW and SDW order, 
       but it weakens the bond-singlet superconducting order.
       Thus, the electric field term is the relevant term for superconductivity, which is consistent with Eq.~(\ref{cpK}).

       We also study the wave vector of the leading superconducting order parameter, which can be determined by Fourier analysis of the corresponding correlation function~\cite{PhysRevLett.105.146403}
       \begin{align}
       	S_{\Phi_{0,0}}(k) := \sum_r e^{-ikr}C_{\Phi_{0,0}}(r).
       \end{align}
       Figure~\ref{fig_6} shows the absolute value of $S_{\Phi_{0,0}}(k)$ at $3/8$ filling.
       For small $h$, there are only two peaks at $k=\pm 2k_F = \pm 3\pi/4$, where $k_F= n\pi $ is the Fermi wave vector with $n$ being the filling factor.
       However, for large $h$, we can find that there is a clear leading peak at $k=\pi$ for $|S_{\Phi_{0,0}}(k)|$, with two subleading peaks at $k=\pm 2k_F$.
       This shows that the  dominant superconducting order for the large $h$ is a PDW with  $\pi$ momentum.
       Therefore, in addition to enhancing the superconducting order, the confinement of lattice fermions can also induce a  $\pi$ momentum.

	\begin{figure}[t] \includegraphics[width=0.45\textwidth]{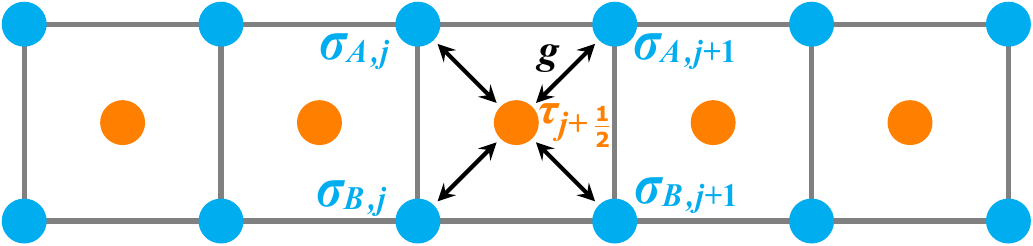} 
	\caption{ Lattice  skeleton  of Eq.~(\ref{Hqs}). Top/bottom blue sites represent $A/B$-spins (matter fields), 
		and  orange sites are $\tau$-spins (gauge fields). Each $\tau$-spin can couple to its 4 NN $\sigma$-spins.}
	\label{fig_7}
\end{figure}

	\section{Proposed experimental implementation.}\label{Sec6}
	
	Finally, we present an approach to realize a 1D $\mathbb{Z}_2$ LGT coupled to two-component fermions in quantum simulators.
	According to the above discussion, we know that $\hat H_U$ is an irrelevant term for low-energy physics.
	Thus, without loss of generality, we just need to consider the Hamiltonian (\ref{H}) with $U=0$ in quantum simulations, 
	which is sufficient to demonstrate the above physics.
	Here we mainly apply an array of spins (or two-level systems) with NN hopping $g$, where the lattice configuration is shown in Fig.~\ref{fig_7}.
	Thus, the original Hamiltonian of this system can be written as
	 \begin{align} \label{Hqs}\nonumber
		\hat  H_{qs}  &= g\sum_{j}\sum_{\ell =A,B} \big( \hat \sigma^+_{\ell, j }\hat\tau^-_{j+\frac{1}{2}}+\hat\sigma^+_{\ell, j+1 }\hat\tau^-_{ j+\frac{1}{2}} + \text{H.c.}\big) \\
		&+\sum_{j} \big( V_{A}\hat \sigma^+_{A, j }\hat\sigma^-_{A, j }+ V_{B}\hat \sigma^+_{B, j }\hat\sigma^-_{B, j }+h\hat\tau^x_{j+\frac{1}{2} }\big).
	\end{align}
	We let the potential of $A$- and $B$-spins satisfy $V_{A}=-V_{B}\gg g,h$.
	Using the Schrieffer-Wolf transformation, we can obtain the effective spin Hamiltonian as
   \begin{align} \label{Hqs1}\nonumber
		\hat  H_{e}  = &\sum_{j} \big( g_e\hat \sigma^+_{A, j }\hat\tau^z_{j+\frac{1}{2}}\hat \sigma^-_{A, j +1}-
		g_e\hat \sigma^+_{B, j }\hat\tau^z_{j+\frac{1}{2}}\hat \sigma^-_{B, j +1} \\
		&+ \text{H.c.}\big) 
		+h\sum_{j} \hat\tau^x_{j+\frac{1}{2}},
	\end{align}
	where $g_e=\lambda^2/V_A$. This is a $\mathbb{Z}_2$ LGT coupled to two species of spins.
	To map the Hamiltonian $\hat H_e$ to Eq.~(\ref{H}), we first apply a particle-hole transformation of the $B$-spin 
	to change the sign of the coupling between $B$-spins and gauge fields.
	Then, via a Jordan-Wigner transformation, we can map $A/B$-spins to spin-$\frac{1}{2}$ fermions,
	and the final Hamiltonian will become  Eq.~(\ref{H}) with $U=0$.
	Detailed derivations can be found in Appendix~\ref{A2}.
	Generally, the Hamiltonian in Eq.~(\ref{Hqs}) is accessible in various of artificial quantum many-body systems,
	including optical lattice~\cite{doi:10.1080/00018730701223200,RevModPhys.80.885,Gross995}, 
	Rydberg atoms~\cite{RevModPhys.82.2313,s41567-019-0733-z}, and superconducting circuits~\cite{you2011atomic,GU20171}.

	\section{Summary and outlook}\label{Sec7}
	We have systematically studied the ground state of a spin-$\frac{1}{2}$ fermion chain coupled to a $\mathbb{Z}_2$ LGT.
	In the $\hat G_j =-1$ sector, the model is equivalent to a 1D Hubbard model coupled to a $\mathbb{Z}_2$ LGT.
	At half filling, the system is a Mott insulator, when at least one of $h$ and $U$ is finite, 
	and the spin sector is an antiferromagnetic Heisenberg model with fractionalized spinon excitations.
	In hole-doped systems, the lattice fermion is confined under nonzero electric fields, leading to the emergence of hole-pair bound states.
	Remarkably, we also demonstrate that this hole pair can enhance the superconducting instability,
	and the superconducting order can even be the dominant order for a suitable filling factor and large applied electric field.
	In addition, the confinement can induce a $\pi$ momentum for the dominant superconducting order parameter leading to a PDW.
	We also propose possible experimental realizations of this model in an array of  two-level systems. 
	Our results demonstrate that the confinement of lattice fermions can enhance a superconducting instability,
    which paves the way for understanding  unconventional superconductors with $\mathbb{Z}_2$ LGTs.
    Meanwhile, our model could also be implemented experimentally in state-of-art quantum simulators~\cite{Buluta108, RevModPhys.86.153,Buluta_2011,Gross995,RevModPhys.80.885,doi:10.1080/00018730701223200,s41567-019-0733-z}.
    
    The Hamiltonian (\ref{H}) is reminiscent of the Holstein-Hubbard model~\cite{PhysRevLett.69.1600,PhysRevB.64.094507,PhysRevLett.125.167001}, 
    which is a typical strongly correlated system with both electron-electron  and electron-phonon interactions.
    Thus, studying the relations between LGTs and Holstein-Hubbard model is a relevant question.
    Another particularly interesting and natural extension of our work would be generalizing our model to 2D~\cite{PhysRevB.37.580,PhysRevB.38.745,PhysRevB.38.2926,PhysRevB.45.9976,
    	PhysRevB.62.7850,PhysRevB.37.7940,PhysRevX.6.041049,Gazit2017,PhysRevLett.127.197004},
    which might be relevant for understanding high-$T_c$ superconductors.
    It will also be an interesting issue to consider a ladder model~\cite{PhysRevResearch.3.013133}, which is one of the simplest cases that open a spin gap~\cite{doi:10.1126/science.271.5249.618,PhysRevB.45.5744,PhysRevLett.91.137203,PhysRevB.85.035104}.
	
	\begin{acknowledgements} 
		We thank M. Dalmonte, E. Rinaldi and R.-Z. Huang for insightful discussions. This work is supported in part by:
		Nippon Telegraph and Telephone Corporation (NTT) Research,
		the Japan Science and Technology Agency (JST) [via
		the Quantum Leap Flagship Program (Q-LEAP), and
		the Moonshot R\&D Grant Number JPMJMS2061],
		the Japan Society for the Promotion of Science (JSPS)
		[via the Grants-in-Aid for Scientific Research (KAKENHI) Grant No. JP20H00134],
		the Asian Office of Aerospace Research and Development (AOARD) (via Grant No. FA2386-20-1-4069), and
		the Foundational Questions Institute Fund (FQXi) via Grant No. FQXi-IAF19-06.
	\end{acknowledgements}

\begin{appendix}
	
\section{Effective Hamiltonian in the case of strong tension}\label{A1}
Here we present details of deriving the effective Hamiltonian with the Schrieffer-Wolf transformation~\cite{PhysRev.149.491,BRAVYI20112793} in the case of $h\gg t$ and $U\rightarrow 0$.
First, we rewrite the Hamiltonian as 
\begin{align} \nonumber
	&\hat  H = \hat H_0 + \hat H_1,\\ \nonumber
	&\hat H_0 = -h\sum_{j} \hat \tau_{j+\frac{1}{2}}^x,\\ 
	&\hat H_1 = -t\sum_{j}\sum_{\sigma=\uparrow,\downarrow}\big(\hat f^\dagger_{j,\sigma}\hat \tau^z_{j+\frac{1}{2}}\hat f_{j+1,\sigma} + \text{H.c.}\big), 
\end{align}
where $\hat H_0$ is a diagonal term, while $\hat H_1$ is off-diagonal one.
Now, for $h\gg t$, we use the Schrieffer-Wolff transformation~\cite{PhysRev.149.491,BRAVYI20112793} to obtain the effective Hamiltonian
\begin{align}
	\hat H_{\text {eff}} = e^{-\hat S}\hat H e^{\hat S}.
\end{align}
To second order:
\begin{align}  \label{hs} \nonumber
	\hat H_{\text {eff}} =  & \hat H_0 + (\hat H_1+[\hat H_0,\hat S])\\
	&+\frac{1}{2}[ (\hat H_1+[\hat H_0,\hat S]),\hat S] 
	+ \frac{1}{2}[\hat H_1,\hat S].
\end{align}
When this condition holds
\begin{align}   \label{Scon}
	\hat H_1+[\hat H_0,\hat S]=0,
\end{align}
then the final effective Hamiltonian reads
\begin{align}  \label{heff}
	\hat H_{\text {eff}} =  \hat H_0 + \frac{1}{2}[\hat H_1,\hat S].
\end{align}
Here we  choose the form of $\hat S$ as
\begin{align}   \label{S}
	\hat S = \frac{it}{2h} \sum_{j,\sigma}\big(\hat f^\dagger_{j,\sigma}\hat \tau^y_{j+\frac{1}{2}}\hat f_{j+1,\sigma} + \text{H.c.}\big),
\end{align}
where it is not difficult to verify that it indeed satisfies the condition in Eq.~(\ref{Scon}).

	\begin{figure*}[t] \includegraphics[width=0.95\textwidth]{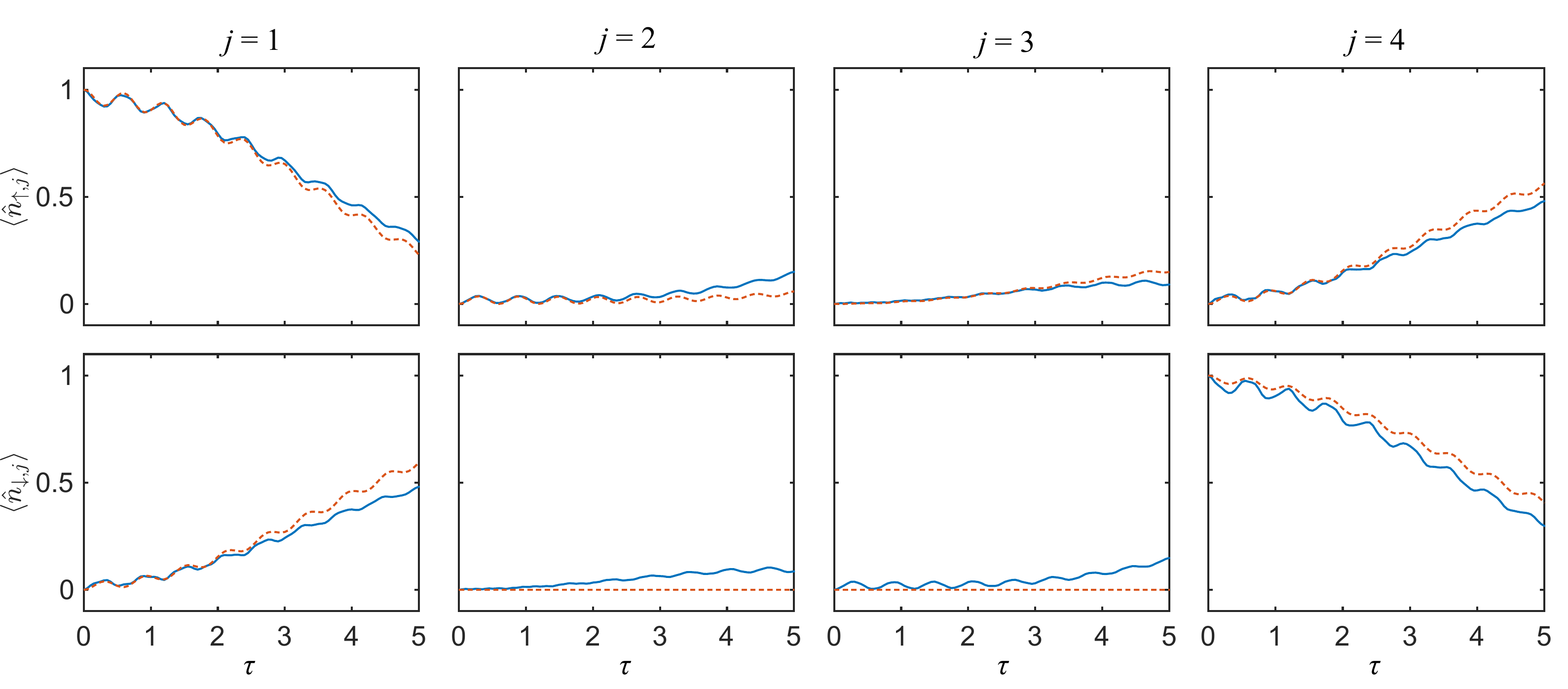} 
	\caption{Quench dynamics of Eqs.~(\ref{H}) and (\ref{hqs})  with $L=4$ (12 qubits).
		The horizontal axis is the time $\tau$.
		The parameters are chosen as $J=20$, $V=400$, $t=g_e=1$, and $h=5$.
		We consider periodic boundary conditions. The initial state is $\ket{\psi_0}=\ket{\uparrow+0-0+\downarrow+}$.
		Solid and dashed curves represent expectation values of the fermion number operator of the Hamiltonians in Eqs.~(\ref{H}) and (\ref{hqs}), respectively. }
	\label{fig_a1}
\end{figure*}
Thus, we have 
\begin{align}  \label{h1s}\nonumber
	[\hat H_1,\hat S] &=-\frac{it^2}{2h} \bigg[ \sum_{i,\sigma'}\big(\hat f^\dagger_{i,\sigma'}\hat \tau^z_{j+\frac{1}{2}}\hat f_{i+1,\sigma'} + \text{H.c.}\big), \\ \nonumber
	&\ \ \ \ \ \ \ \ \ \ \ \ \ \ \ \sum_{j,\sigma}\big(\hat f^\dagger_{j,\sigma}\hat \tau^y_{j+\frac{1}{2}}\hat f_{j+1,\sigma} + \text{H.c.}\big)\bigg] \\ \nonumber
	&= -\frac{it^2}{2h} \bigg(  \sum_{j,\sigma} \big(2i\hat \tau^x_{j+\frac{1}{2}}\big)\big(2\hat n_{j,\sigma}\hat n_{j+1,\sigma}-\hat n_{j,\sigma}-\hat n_{j+1,\sigma}\big)\\ \nonumber
	& \ \  \ +\sum_{j} \big(-4i\hat \tau^x_{j+\frac{1}{2}}\big)\big(\hat f^\dagger_{j,\uparrow}\hat f^\dagger_{j,\downarrow}\hat f_{j+1,\downarrow}\hat f_{j+1,\uparrow}
	+ \text{H.c.}\big)\\ \nonumber
	& \ \   \ +\sum_{j} \big(4i\hat \tau^x_{j+\frac{1}{2}}\big)\big(\hat f^\dagger_{j,\uparrow}\hat f_{j,\downarrow}\hat f^\dagger_{j+1,\downarrow}\hat f_{j+1,\uparrow}
	+ \text{H.c.}\big)\\
	&\ \ \ \  + \sum_{j,\sigma} \big(\hat \tau^z_{j-\frac{1}{2}}\hat \tau^y_{j+\frac{1}{2}}-\hat \tau^y_{j-\frac{1}{2}}\hat \tau^z_{j+\frac{1}{2}}\big)\big(\hat f^\dagger_{j-1,\sigma}\hat f_{j+1,\sigma} - \text{H.c.}\big)
	\bigg).
\end{align}
We  define spin operators as 
\begin{align}   \label{spin} \nonumber
	&\hat{\boldsymbol{s}}_{j} := \sum_{\alpha,\beta}\hat f^\dagger_{j,\alpha} {\boldsymbol{\sigma}}_{\alpha\beta} \hat f_{j,\beta},\\
	&\hat{\boldsymbol{\eta}}_j := \hat P^{-1} \hat{\boldsymbol{s}}_j\hat P,
\end{align}
where $\hat P$ corresponds to the particle-hole transformation satisfying
	\begin{align} \nonumber
		& \hat P^{-1} \hat f_{j,\uparrow} \hat P =\hat f_{j,\uparrow},\\
		& \hat P^{-1} \hat f_{j,\downarrow} \hat P =(-1)^j \hat f^\dagger_{j,\downarrow}.
\end{align}
Hence, we have 
\begin{align}  \nonumber
	&\hat f^\dagger_{j,\uparrow}\hat f_{j,\downarrow}\hat f^\dagger_{j+1,\downarrow}\hat f_{j+1,\uparrow} + \text{H.c.}
	= \hat s_j^+\hat s_{j+1}^- + \text{H.c.}\\ \nonumber
	&\hat f^\dagger_{j,\uparrow}\hat f^\dagger_{j,\downarrow}\hat f_{j+1,\downarrow}\hat f_{j+1,\uparrow} + \text{H.c.}
	=- (\hat \eta_j^+\hat \eta_{j+1}^- + \text{H.c.})\\
	&\sum_{\sigma}2\hat n_{j,\sigma}\hat n_{j+1,\sigma}-\hat n_{j,\sigma}-\hat n_{j+1,\sigma}
	= \hat s_j^z\hat s_{j+1}^z+ \hat \eta_j^z\hat \eta_{j+1}^z.
\end{align}
Therefore, Eq.~(\ref{h1s}) can be rewritten as 
\begin{align}  \label{h1s2}\nonumber
	[\hat H_1,& \hat S] =\frac{t^2}{h} \sum_{j} \hat \tau^x_{j+\frac{1}{2}}\cdot\big( \hat{\boldsymbol{s}}_j\cdot\hat{\boldsymbol{s}}_{j+1} + \hat{\boldsymbol{\eta}}_j\cdot\hat{\boldsymbol{\eta}}_{j+1} \big)\\ 
	&+ \frac{-it^2}{2h} \sum_{j,\sigma} \big(\hat \tau^z_{j-\frac{1}{2}}\hat \tau^y_{j+\frac{1}{2}}-\hat \tau^y_{j-\frac{1}{2}}\hat \tau^z_{j+\frac{1}{2}}\big)\big(\hat f^\dagger_{j-1,\sigma}\hat f_{j+1,\sigma} - \text{H.c.}\big).
\end{align}
According to Eq.~(\ref{heff}), we can obtain the final form of the effective Hamiltonian 
\begin{align} \label{heff_f}\nonumber
	\hat  H_{\text{eff}} &=	 -h\sum_{j} \hat \tau_{j+\frac{1}{2}}^x
	+J\sum_{j}\hat \tau^x_{j+\frac{1}{2}}\cdot\big( \hat{\boldsymbol{s}}_j\cdot\hat{\boldsymbol{s}}_{j+1} + \hat{\boldsymbol{\eta}}_j\cdot\hat{\boldsymbol{\eta}}_{j+1} \big) \\ 
	&-\frac{iJ}{2}\sum_{j,\sigma}\big(\hat \tau^z_{j-\frac{1}{2}}\hat \tau^y_{j+\frac{1}{2}}-\hat \tau^y_{j-\frac{1}{2}}\hat \tau^z_{j+\frac{1}{2}}\big)\big(\hat f^\dagger_{j-1,\sigma}\hat f_{j+1,\sigma} - \text{H.c.}\big),
\end{align}
where $J=t^2/h$. Using the identity 
\begin{align} 
\hat \tau^z_{j-\frac{1}{2}}\hat \tau^y_{j+\frac{1}{2}}-\hat \tau^y_{j-\frac{1}{2}}\hat \tau^z_{j+\frac{1}{2}} = i\hat\tau^x_{j+\frac{1}{2}}\cdot(\hat \tau^z_{j-\frac{1}{2}}\hat \tau^z_{j+\frac{1}{2}}\!+\!\hat \tau^y_{j-\frac{1}{2}}\hat \tau^y_{j+\frac{1}{2}}),
\end{align}
we can finally obtain the effective Hamiltonian in Eq.~(\ref{Heff}).

\section{Proposed experimental implementation}\label{A2}
Here we show details about how to realize the Hamiltonian (\ref{H}) in quantum simulators.
We consider spin (or two-level) systems with spin exchange coupling, as well as tunable longitudinal and transverse fields.
The lattice configuration is shown in Fig.~\ref{fig_7},
and the original Hamiltonian of this system can be written as
\begin{align} \label{hqs}\nonumber
	&\hat  H_{qs}  = \hat  H_{qs,0}+\hat  H_{qs,1},\\ \nonumber
	&\hat  H_{qs,0}=\sum_{j}V_{A}\hat \sigma^+_{A, j }\hat\sigma^-_{A, j }+ V_{B}\hat \sigma^+_{B, j }\hat\sigma^-_{A, j }-h\hat\tau^x_{j+\frac{1}{2} },\\
	&\hat  H_{qs,1} = g\sum_{j}\sum_{\ell =A,B} \hat \sigma^+_{\ell, j }\hat\tau^-_{j+\frac{1}{2}}+\hat\sigma^+_{\ell, j+1 }\hat\tau^-_{ j+\frac{1}{2}} + \text{H.c.}.
\end{align}
To realize a $\mathbb{Z}_2$ LGT coupled to a two-component matter field, we let $V_{A}=-V_{B} =V\gg g,h$.

\begin{figure*}[t] \includegraphics[width=1\textwidth]{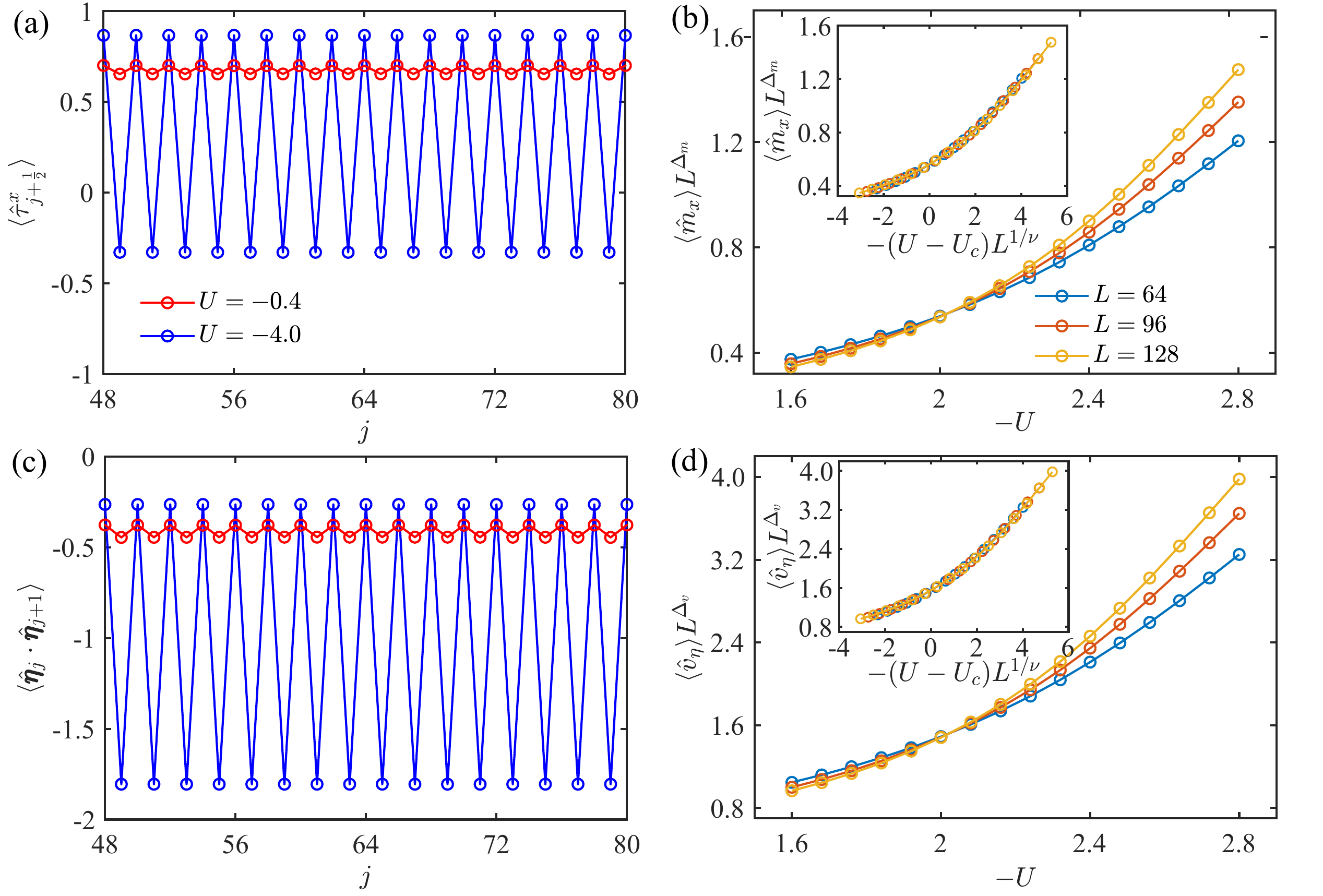} 
	\caption{ Quantum phase transition for negative $U$ and $h=1$ at half filling.
	(a) Expectation value distributions of $\hat\tau^x_{j+\frac{1}{2}}$ at the ground state for small and large $-U$.
     (b) Rescaled order parameter $\braket{\hat m_x}L^{\Delta_m}$ as a function of $-U$ for different system sizes, where the dimension $\Delta_m\approx0.33$. 
     The curves cross at the critical point $-U_c\approx2.04$.
     The insert shows the corresponding data collapse with a correlation length critical exponent $\nu\approx2.5$.
 (c) Expectation value distributions of $\hat{\boldsymbol{\eta}}_{j}\cdot\hat{\boldsymbol{\eta}}_{j+1}$ at the ground state for small and large $-U$.
 (d) Rescaled order parameter $\braket{\hat v_\eta}L^{\Delta_v}$ as a function of $-U$ for different system sizes, where the dimension $\Delta_v\approx\Delta_m\approx0.33$. 
 The curves  also cross at the critical point.
 The insert shows the corresponding data collapse.}
	\label{fig_a2}
\end{figure*}

We apply the Schrieffer-Wolf transformation to obtain the effective spin Hamiltonian.
Similar to Ref.~\cite{Ge_2021}, the generating function can be chosen as
\begin{align} \nonumber
	\hat S'  = \frac{ig}{V}\sum_{j }  & \hat \sigma^+_{A, j }\hat\tau^-_{j+\frac{1}{2}}+\hat\sigma^+_{A, j+1 }\hat\tau^-_{ j+\frac{1}{2}} \\
	&- \hat \sigma^+_{B, j }\hat\tau^-_{j+\frac{1}{2}}-\hat\sigma^+_{B, j+1 }\hat\tau^-_{ j+\frac{1}{2}} -\text{H.c.}.
\end{align}
According to Eq.~(\ref{heff}), we can obtain the final effective Hamiltonian of $\hat H_{qs}$ as
\begin{align} \label{hqs1}\nonumber
	\hat  H_{e}  = &\hat H_{qs,0} + \frac{1}{2}[\hat H_{qs,1},\hat S']\\ \nonumber
	=&g_e\sum_{j} \big( \hat \sigma^+_{A, j }\hat\tau^z_{j+\frac{1}{2}}\hat \sigma^-_{A, j +1}-
	\hat \sigma^+_{B, j }\hat\tau^z_{j+\frac{1}{2}}\hat \sigma^-_{B, j +1} + \text{H.c.}\big) \\ \nonumber
	&+g_e\sum_{j} \big [ \hat \tau^+_{ j -\frac{1}{2}} (\hat\sigma^z_{B,j}-\hat\sigma^z_{A,j})\hat \tau^-_{ j +\frac{1}{2}}+ \text{H.c.}\big] \\
	&+\sum_{j} (V+2g_e)\hat \sigma^+_{A, j }\hat\sigma^-_{A, j }+ (V-2g_e)\hat \sigma^+_{B, j }\hat\sigma^-_{B, j } -h\hat\tau^x_{j+\frac{1}{2}},
\end{align}
where $g_e = g^2/V$.
Here we can find that the total spins of  the $A$ and $B$ sublattices are both conserved, i.e., $[\sum_{j} \hat \sigma^+_{\ell, j }\hat\sigma^-_{\ell, j }, \hat H_e]=0$.
Thus, the potential terms of $A/B$ spins can be neglected.
In addition, for large $h$, the second term of Eq.~(\ref{hqs1}) is irrelevant~\cite{Ge_2021}.
Therefore, the effective Hamiltonian in this case can be simplified as
\begin{align} \label{he}\nonumber
	\hat  H_{e}  =&g_e\sum_{j} \big(  \hat \sigma^+_{A, j }\hat\tau^z_{j+\frac{1}{2}}\hat \sigma^-_{A, j +1}-
	\hat \sigma^+_{B, j }\hat\tau^z_{j+\frac{1}{2}}\hat \sigma^-_{B, j +1}+ \text{H.c.}\big) \\
	& -h\sum_{j}\hat\tau^x_{j+\frac{1}{2}}.
\end{align}
Now we apply a particle-hole transformation of $B$-sites $\hat P_B = \prod_{j=\text{odd}}\hat\sigma^z_{B, j }$, 
which can change the sign of the coupling strength between $B$- and $\tau$-spins.
That is, 
\begin{align} \nonumber
	\hat  H_{e} \rightarrow  \hat P_B\hat  H_{e}\hat P_B  =g_e\sum_{j} \big( & \hat \sigma^+_{A, j }\hat\tau^z_{j+\frac{1}{2}}\hat \sigma^-_{A, j +1}+
	\hat \sigma^+_{B, j }\hat\tau^z_{j+\frac{1}{2}}\hat \sigma^-_{B, j +1} \\
	&+ \text{H.c.}\big) 
	-h\sum_{j}\hat\tau^x_{j+\frac{1}{2}}.
\end{align}
Finally,  to map the matter field from spins to  fermions, we can use a Jordan-Wigner transformation defined as
\begin{align} \nonumber
	&\hat f^\dagger_{j,\uparrow} = \hat \sigma^+_{A, j}\prod_{k<j}\hat\sigma^z_{A, k },\\
	&\hat f^\dagger_{j,\downarrow} = \hat \sigma^+_{B, j}\prod_{k<j}\hat\sigma^z_{B, k }\prod_{l=1}^L\hat\sigma^z_{A, l }.
\end{align}
Hence, $\hat  H_{e} $ can be mapped to the Hamiltonian in Eq.~(\ref{H}), i.e., a spin-$\frac{1}{2}$ fermion chain minimally coupled to a $\mathbb{Z}_2$ LGT.
We also perform a numerical simulation to further demonstrate the above discussion.
Here we use exact diagonalization to calculate the quench dynamics of Eqs.~(\ref{H}) and (\ref{hqs})  with $L=4$ (12 qubits).
The parameters chosen are $J=20$, $V=400$, $t=g_e=1$, and $h=5$.
To reduce the boundary effect, we consider periodic boundary conditions.
The initial state is $\ket{\psi_0}=\ket{\uparrow+0-0+\downarrow+}$.
As shown in Fig.~\ref{fig_a1}, we can find consistent fermion density results when comparing the dynamics of  
Eqs.~(\ref{H}) and (\ref{hqs}).

\section{Negative $U$ at half filling}\label{A3}
In this appendix, we discuss the phase diagram of the Hamiltonian (\ref{H}) with negative $U$ at half filling.
In the gauge sector $\hat G_j = -1$, negative $U$ means antiferromagnetic interactions of electric fields or attractive on-site interactions of fermions.
As discussed in Sec.~\ref{Sec3}, due to the confinement of lattice fermions, exciting a double occupation and a hole (called meson) with distance $r$ costs energy $\sim (hr+U)$.
When $-U\ll h$, this excitation should be absent in the ground state due to the large energy cost.
Thus, the gauge field tends to polarize at $\hat\tau_j^x=1$,
and  the lattice fermion prefers the single-occupation state, i.e., the system should be a Mott insulator with gapless spin sector.
However, when $ -U\gg h$, the energy cost of exciting a meson is negative, 
so double occupations and holes tend to dominate resulting in the freeze of the spin sector, i.e., opening a spin gap.
Meanwhile, in this case, $\hat \tau^x$ tends to have a staggered distribution to make $\braket{\hat\tau_{j-\frac{1}{2}}^x\hat\tau_{j+\frac{1}{2}}^x}<0$.
Therefore, for fixed $h$,  when increasing $-U$, there should be a quantum phase transition 
from the Mott insulator (gapless) to meson condensed phase (gaped).

For large $-U$, since $\hat \tau^x$ is expected to have a staggered distribution, 
the ground state should be dimerized, see Fig.~\ref{fig_a2}(a).
Thus, the condensation of mesons induces a $\pi$ momentum, i.e., a spontaneous breaking of the translational symmetry occurs.
To study this quantum phase transition, we can choose the order parameter
\begin{align}
\hat m_x := \frac{2}{N}\sum_{j}(\hat\tau_{2j-\frac{1}{2}}^x-\hat\tau_{2j+\frac{1}{2}}^x).
\end{align}
For numerical simulations, to reduce the boundary effect, 
we obtain the order parameter as $\hat m_x = \hat\tau_{L/2-\frac{1}{2}}^x-\hat\tau_{L/2+\frac{1}{2}}^x$.
In Fig.~\ref{fig_a2}(b), we present the expectation values of $\hat m_x$ versus $-U$ for $h=1$ and different system sizes.
The numerical result show that the critical point is at $-U_c \approx 2.04$ with critical exponents $\Delta_m\approx0.33$ and $\nu\approx2.5$, 
where $\Delta_m$ and $\nu$ are  the order parameter dimension and correlation length critical exponent, respectively.

To further understand this quantum phase transition, we now focus on the matter field.
When $h=0$, for large $-U$, the lattice fermion at half filling has a gaped spin sector and a gapless charge sector~\cite{Giamarchi2004,Fradkin2013}.
However, according to the above discussion, we find that the charge sector becomes gaped under finite $h$  for large $-U$.
In addition, due to the pseudospin rotation symmetry ($SU_{\eta}(2)$), 
which cannot be broken spontaneously in 1D systems, 
this gaped translational symmetry breaking phase cannot be a CDW.
Therefore, to preserve the $SU_{\eta}(2)$ symmetry, we expect that the charge sector is a pseudospin valence-bond solid, see Fig.~\ref{fig_a2}(c).
This order can be described by the order parameter
\begin{align}
\hat v_\eta := \frac{2}{N}\sum_{j}(\hat{\boldsymbol{\eta}}_{2j-1}\cdot\hat{\boldsymbol{\eta}}_{2j}-\hat{\boldsymbol{\eta}}_{2j}\cdot\hat{\boldsymbol{\eta}}_{2j+1}).
\end{align}
In Fig.~\ref{fig_a2}(d), we present the expectation values of $\hat v_\eta$ versus $-U$ for $h=1$ and different system sizes, 
where we calculate it as $\hat v_\eta=\hat{\boldsymbol{\eta}}_{L/2-1}\cdot\hat{\boldsymbol{\eta}}_{L/2}-\hat{\boldsymbol{\eta}}_{L/2}\cdot\hat{\boldsymbol{\eta}}_{L/2+1}$ to reduce the boundary effect.
We can find that $\hat v_\eta $ can also represent this quantum phase transition and has the same dimension as $\hat m_x$ at the critical point.

\end{appendix}

%


\end{document}